\documentclass[jkps,preprint,fleqn,showpacs,showkeys]{revtex4}
\usepackage{graphicx,color}
\usepackage{amssymb}
\usepackage{amsmath}
\usepackage{bm}
\begin{document}
\setcounter{page}{0}
\title[]{Correlation between Ultra-high Energy Cosmic Rays and Active Galactic Nuclei from Fermi Large Area Telescope}
\author{Jihyun \surname{Kim}}
\email{jihyunkim@hanyang.ac.kr}
\affiliation{Department of Physics, Hanyang University, Seoul
133-791}
\author{Hang Bae \surname{Kim}}
\email{hbkim@hanyang.ac.kr}
\affiliation{Department of Physics, Hanyang University, Seoul
133-791}

\begin{abstract}
We study the possibility that the $\gamma$-ray loud active
galactic nuclei (AGN) are the sources of ultra-high energy
cosmic rays (UHECR), through the correlation analysis of
their locations and the arrival directions of UHECR. We use
the $\gamma$-ray loud AGN with $d\le\,100\,{\rm Mpc}$ from
the second Fermi Large Area Telescope AGN catalog and the
UHECR data with $E\ge\,55\,{\rm EeV}$ observed by Pierre
Auger Observatory. The distribution of arrival directions
expected from the $\gamma$-ray loud AGN is compared with
that of the observed UHECR using the correlational
angular distance distribution and the Kolmogorov-Smirnov test.
We conclude that the hypothesis that the $\gamma$-ray loud
AGN are the dominant sources of UHECR is disfavored unless
there is a large smearing effect due to the intergalactic
magnetic fields.
\end{abstract}



\maketitle

\section{Introduction}
The origin of ultra-high energy cosmic rays (UHECR), whose energies
are above $1\,{\rm EeV}(=10^{18}\,{\rm eV})$, has been searched
for many years; however, it is still in vague. In the search
for the origin of UHECR, the Greisen-Zatsepin-Kuzmin (GZK) suppression
\cite{Greisen:1966jv, Zatsepin:1966jv} plays an important role.
This suppression tells us that the sources of UHECR with energies
above the GZK cutoff, $E_{\rm GZK}\sim 40\,{\rm EeV}$, should
be located within the GZK radius, $r_{\rm GZK}\sim100\,{\rm Mpc}$,
because the UHECR coming from beyond the GZK radius cannot reach
us by loosing energies as consequences of the interactions between
cosmic microwave background photons. The recent observations
\cite{Abbasi:2007sv, Abraham:2010mj, Tsunesada:2011mp} support
the GZK suppression, thus we can focus on the source candidates
lying within $\sim100\,{\rm Mpc}$.

For the possible sources of UHECR, several kinds of astrophysical
objects have been proposed, such as $\gamma$-ray bursts,
radio galaxies, and active galactic nuclei (AGN) \cite{Hillas:1985is,
Waxman:1995dg, Vietri:1995hs, Rachen:1992pg, Halzen:1997hw,
Milgrom:1995um,Waxman:1995vg,Stanev:1996qc} which are known to be
able to accelerate UHECR enough. To verify that these astrophysical
objects could be the sources of UHECR, it is worthwhile to
compare the arrival directions of UHECR and the positions of
source candidates using statistical tests. Although many statistical
studies for correlation have been done \cite{Torres:2002bb, Singh:2003xr,
Gorbunov:2004bs, Abbasi:2005qy, Abbasi:2008md, Cronin:2007zz, :2010zzj,
Kim:2010zb, Kim:2012en}, the origin of UHECR is not confirmed yet.

In our previous work \cite{Kim:2010zb, Kim:2012en}, we examined
the AGN model, where UHECR with energies above a certain energy
cutoff are generated from the AGN lying within a certain distance
cut. We used AGN listed in V\'eron-Cetty and V\'eron (VCV) catalog
\cite{VCV:12, VCV:13} and the UHECR data observed by Pierre Auger
Observatory (PAO) \cite{Cronin:2007zz, :2010zzj}. By statistical
test methods, we concluded that the whole AGN listed in VCV catalog
cannot be the true sources of UHECR and pointed out that a certain
subset of listed AGN could be the true sources of UHECR. Some subsets
of AGN, which have the counterpart in X-ray or $\gamma$-ray bands
have been tested for the possibility that they are responsible
for UHECR \cite{:2010zzj, Harari:2008zp, George:2008zd, Nemmen:2010bp,
Jiang:2010yc}. In the case of the correlation with AGN detected in
hard X-ray band, it is found that the fractional excess of pairs
relative to the isotropic expectation \cite{:2010zzj}. In the case of AGN detected
in $\gamma$-ray band, a marginal correlation is found when we consider
the small angular separations only, but the strong correlation is found
when we consider the angular separations up to $20^{\circ}$ \cite{Nemmen:2010bp}.

This paper focuses on the statistical tests for the correlation of
UHECR with certain subsets of AGN, especially $\gamma$-ray loud
AGN because $\gamma$-ray loud AGN have sufficient power to accelerate UHECR.
In Section \ref{sec2}, the source models of UHECR which
assume that UHECR with $E\ge E_c$ come from AGN with $d\le d_c$ which
are emitting strong $\gamma$-rays is presented. Based on the
$\gamma$-ray loud AGN model we construct, we create the mock arrival
directions of UHECR by Monte-Carlo simulation.
In Section \ref{sec3}, we describe the UHECR data and the $\gamma$-ray
loud AGN data we use in our analysis. We use 2010 PAO data
\cite{:2010zzj} for observed UHECR data and the second catalog of
AGN detected by the Fermi Large Area Telescope (LAT) \cite{Fermi:2011bg}
for the $\gamma$-ray loud AGN data. To test the correlation between
the observed UHECR and the $\gamma$-ray loud AGN, the methods which
can compare the distribution of observed UHECR arrival direction and
that of the mock UHECR arrival direction expected from the source model
are needed to be established. The brief descriptions of our test methods
are provided in Section \ref{sec4}. The results of statistical tests
are given in Section \ref{sec5} and the discussion and the conclusion follow
in Section \ref{sec6}.

\section{Source model of UHECR}
\label{sec2}
Several astrophysical objects are know to be able to accelerate
CR up to ultra-high energy. Among them, AGN are the most popular objects
since the strong correlation was claimed by PAO \cite{Cronin:2007zz}.
However, its updated analysis results and other studies exclude the
hypothesis that the whole set of AGN is responsible for the UHECR
\cite{:2010zzj, Abbasi:2008md, Kim:2010zb, Kim:2012en}. In our previous
work \cite{Kim:2012en}, using PAO UHECR having energies above $55\,{\rm EeV}$
\cite{:2010zzj} and AGN within $100\,{\rm Mpc}$ listed in the 13th
edition of VCV catalog \cite{VCV:13}, we concluded that we can reject
the hypothesis that the whole AGN within $100\,{\rm Mpc}$ are the real sources
of UHECR. Also, we tested the possibility that the subset of AGN
is responsible for UHECR; we took AGN within arbitrary distance band as source
candidates and found a good correlation for AGN within $60-80\,{\rm Mpc}$.
However, we do not have a reasonable physical explanation for
distance grouping. This motivates us to try the subclass of AGN with
proper physical properties appropriate for UHECR acceleration.

In this paper, we study the hypotheses that AGN emitting strong
$\gamma$-ray are the sources of UHECR, based on the theoretical
study in Ref. \cite{Dermer:2010iz}. Dermer et al. calculated the
emissivity of non-thermal radiation from AGN using the first Fermi
LAT AGN catalog (1LAC) \cite{Fermi:2010ge} to confirm that they
have sufficient power to accelerate UHECR. In the Fermi acceleration
mechanism using colliding shell model, they found that some of AGN
listed in 1LAC have enough power to accelerate UHECR.
(See the Figure 3. in \cite{Dermer:2010iz}.)
Therefore, we set up the $\gamma$-ray loud AGN model
for the UHECR source in the same way as our AGN model introduced
in our previous work \cite{Kim:2012en}.

When UHECR propagate through the universe, UHECR undergo
deflection of its trajectory by intergalactic magnetic fields.
These phenomena are embedded in the simulation for the mock
UHECR expected from the AGN model to compare with observed UHECR
by introducing the smearing angle parameter ($\theta_s$) and
by restricting the distance of the source ($d_c$) and the
energy of observed UHECR ($E_c$) following the GZK suppression.

We study two versions of $\gamma$-ray loud AGN models in this
paper. The first one assumes that UHECR with energies
$E\ge E_c$ come from AGN which are emitting strong
$\gamma$-rays with distance $d\le d_c$, and the second one
assumes that among the $\gamma$-ray loud AGN those having
TeV or very high energies $\gamma$-ray emission are responsible
for the UHECR. The same constraints of UHECR energy and AGN
distance are applied to these models. We call the first one
the $\gamma$-ray loud AGN model, and the second one the TeV
$\gamma$-ray AGN model.

The UHECR flux in the simulation for these two models
are described below. The expected UHECR flux
at a given arrival direction $\hat{\bf r}$ is
composed of the $\gamma$-ray loud AGN contribution
and the isotropic background contribution
\begin{equation}
F(\hat{\bf r}) = F_{\rm AGN}(\hat{\bf r}) + F_{\rm ISO},
\end{equation}
where $F_{\rm AGN}(\hat{\bf r})$ is the distribution of all AGN
within the distance cut and $F_{\rm ISO}$ the contribution of
isotropic background from the outside of distance cut $d_c$.
That is, a certain fraction of UHECR is coming from AGN
and the remaining fraction of them is originated from
the isotropic background. We introduce the AGN fraction
parameter $f_A$ as
\begin{equation}
f_A = \frac{\overline{F}_{\rm AGN}}{\overline{F}_{\rm AGN}+F_{\rm ISO}},
\end{equation}
where $\overline{F}_{\rm AGN}=(4\pi)^{-1}\int F_{\rm AGN}(\hat{\bf r})d\Omega$
is the average AGN-contributed flux.

In the next step, we consider two approaches for $F_{\rm AGN}(\hat{\bf r})$
because the relation between UHECR flux and AGN property is not
established yet. The UHECR flux from all AGN can be written as
\begin{equation}
F_{\rm AGN}(\hat{\bf r})\propto\sum_{j\in{\rm AGN}}
\frac{L_j}{4\pi d_j^2}\cdot
\exp\left[-\left(\theta_j(\hat{\bf r})/\theta_{sj}\right)^2\right]\,,\\
\label{UHECR-flux}
\end{equation}
where $L_j$ is the UHECR luminosity, $d_i$ is the distance,
$\theta_j(\hat{\bf r})=\cos^{-1}(\hat{\bf r}\cdot\hat{\bf r}'_j)$
is the angle between the direction $\hat{\bf r}$ and the $j$-th AGN,
$\theta_{sj}$ is the smearing angle of the $j$-th AGN. The first
approach assumes that all AGN have the same UHECR luminosity,
$L_j=L$, and the same smearing angle, $\theta_{sj}=\theta_s$.
The second approach assumes that the UHECR flux contributed by
AGN is proportional to the $\gamma$-ray flux of AGN.
\begin{equation}
F_{\rm AGN}(\hat{\bf r})\propto\sum_{j\in{\rm AGN}}
F_{\gamma,j}\cdot
\exp\left[-\left(\theta_j(\hat{\bf r})/\theta_{sj}\right)^2\right]\,,\\
\label{UHECR-photon-flux}
\end{equation}
where $F_{\gamma,j}$ is the photon flux of AGN detected by Fermi LAT
in the $1-100\,{\rm GeV}$ energy band. Although the normalization
is needed for the accurate expression of Eq. (\ref{UHECR-flux}) and
(\ref{UHECR-photon-flux}), we neglect it because we are not concerned
with the total flux of UHECR in the test.

We have two free parameters in our model for the simulation,
the smearing angle $\theta_s$ and the AGN fraction $f_A$.
For the fiducial values of $\theta_s$ and $f_A$, we take
$\theta_s=6^{\circ}$ \cite{Kashti:2008bw} and $f_A=0.7$ \cite{Koers:2008ba}.

In the last step for realizing the mock UHECR in the simulation,
we need to consider the exposure function reflecting the
efficiency of the detector. The geometric efficiency of the detector
depends on the location of the experimental site and
the zenith angle cut.
Then, the exposure function $h(\delta)$ is given by \cite{Sommers:2000us}
\begin{equation}
h(\delta) = \frac{1}{\pi}\left[ \sin\alpha_m\cos\lambda\cos\delta
	+\alpha_m\sin\lambda\sin\delta\right],
\end{equation}
where $\lambda$ is the latitude of the detector array,
$\theta_m$ is the zenith angle cut, and
\[
\alpha_m=\left\{\begin{array}{ll}
0,            & \hbox{for\ } \xi > 1, \\
\pi,          & \hbox{for\ } \xi < -1, \\
\cos^{-1}\xi, & \hbox{otherwise}
\end{array}\right.
\ \hbox{with}\
\xi=\frac{\cos\theta_m-\sin\lambda\sin\delta}{\cos\lambda\cos\delta}.
\]
The latitude of the PAO site is $\lambda=-35.20^\circ$
and the zenith angle cut of the released data is $\theta_m=60^\circ$.

\section{Description of the data}
\label{sec3}
We get the information on the $\gamma$-ray loud AGN from
the second catalog of Fermi LAT AGN (2LAC) published in
2011 \cite{Fermi:2011bg}. The 2LAC includes the AGN information
collected by the Fermi LAT for two years. It contains
1017 $\gamma$-ray sources located at high galactic latitude
($|b|>10^{\circ}$) as the low galactic latitude region is masked
by the galactic plane.That region is excluded
because the low galactic latitude region is too noisy to be
investigated due to diffuse radio emission, interloping
galactic point sources, and heavy optical extinction.
There are 886 AGN samples, which are called clean AGN,
categorized by the condition that the sole AGN is
associated with the $\gamma$-ray source and has the
association probability $P$ is larger than $0.8$.

In the $\gamma$-ray loud AGN model, we pick up the distance cut
$d_c=100\,{\rm Mpc}$ corresponding to the redshift $z\sim0.024$.
(We use $h=0.70$, $\Omega_m=0.27$, and $\Omega_{\Lambda}=0.73$
to convert the redshift to the distance.)
Then, only 8 AGN among the clean AGN are picked up as UHECR source candidates.
For the TeV AGN model, we use the list of TeV AGN detected
by the Fermi LAT. There are 34 TeV AGN among the clean AGN
and only 3 TeV AGN are used for the source candidate after the distance cut.
(See the Table 9 in \cite{Fermi:2011bg}.)
In the Table \ref{list}, we list 8 source candidate AGN within
$100\,{\rm Mpc}$, which are detected in $\gamma$-ray range,
and the TeV AGN are marked using TeV flag, Y.

\begin{figure}
    \begin{tabular}{cc}
    \includegraphics[width=0.8\textwidth]{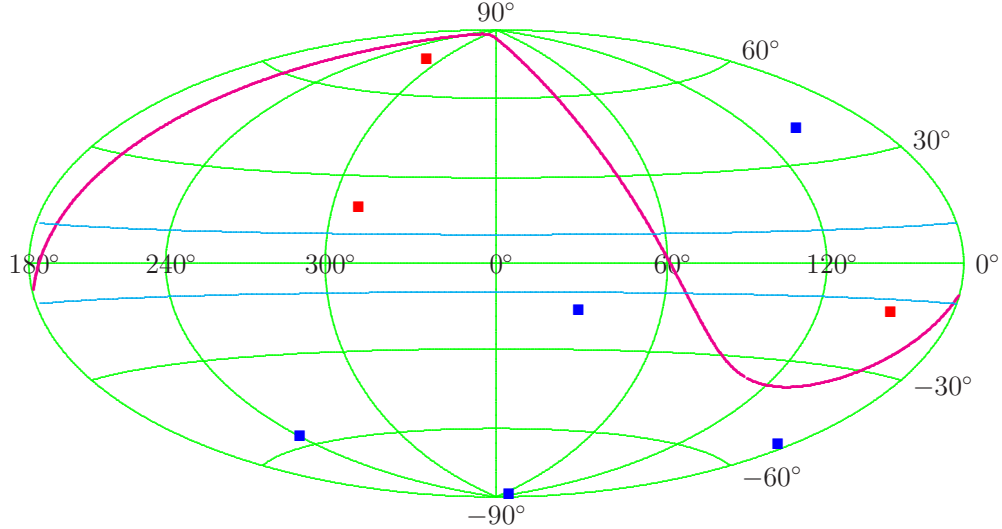}
    \end{tabular}
    \caption{
    Distributions of 8 Fermi LAT AGN within $100\,{\rm Mpc}$ (blue and red squares) and
    57 PAO UHECR (black bullets) in galactic coordinate using the Hammer projection.
    The red squares represent $\gamma$-ray emitting AGN in TeV band. The magenta line
    means a boundary of PAO field of view and the cyan lines represent the border of
    the low latitude region ($|b|<10^{\circ}$).
    }
\label{skymap}
\end{figure}

We use PAO 2010 data \cite{:2010zzj} for the observed UHECR data,
which were collected by the surface detector from 2004-01-01 to
2009-12-31. It includes 69 events in the declination band
$\delta=-90^{\circ}\textrm{--}24.8^{\circ}$ in the equatorial
coordinates and having energies above $55\,{\rm EeV}$. Among them,
we use only 57 PAO UHECR to avoid the galactic plane region
$|b|<10^{\circ}$ as mentioned above. Fig.~\ref{skymap} shows
the distributions of the arrival directions of PAO data and
$\gamma$-ray loud AGN detected by Fermi LAT in the galactic
coordinates using the Hammer projection.

\begin{table}
\begin{tabular}{|c|c|c|c|c|c|c|}\hline
Name & $l$ & $b$ & $z$ & $F_{\gamma}$ & \;TeV flag\, & class\\
\hline
\,Centaurus A & ~309.52~ & ~19.42~ & ~0.0008~ & ~$3.03\times 10^{-9}$~ & Y & \;Radio Galaxy\\
NGC 0253 & ~97.37~ & ~-87.96~ & ~0.0010~ & ~$6.2\times 10^{-10}$~ & N & \;Starburst Galaxy\\
M82  & ~141.41~ & ~40.57~ & ~0.0012~ & ~$1.02\times 10^{-9}$~ & N & \;Starburst Galaxy\\
M87 & ~283.78~ & ~74.49~ & ~0.0036~ & ~$1.73\times 10^{-9}$~ & Y & \;Radio Galaxy\\
NGC 1068 & ~172.10~ & ~-51.93~ & ~0.0042~ & ~$5.1\times 10^{-10}$~ & N & \;Starburst Galaxy\\
Fornax A & ~240.16~ & ~-56.69~ & ~0.0050~ & ~$5.3\times 10^{-10}$~ & N & \;Radio Galaxy\\
NGC 6814 & ~29.35~ & ~-16.01~ & ~0.0052~ & ~$6.8\times 10^{-10}$~ & N & \;Unidentified\\
NGC 1275 & ~150.58~ & ~-13.26~ & ~0.018~ & ~$1.88\times 10^{-8}$~ & Y & \;Radio Galaxy\\
\hline
\end{tabular}
\caption{The 8 clean $\gamma$-ray loud AGN within $100\,{\rm Mpc}$.
$l$: galactic longitude (degrees),
$b$: galactic latitude (degrees),
$z$: redshift,
$F_{\gamma}$: photon flux (${\rm photon/cm^2/s}$),
TeV flag: TeV AGN,
class: optical class}
\label{list}
\end{table}

\section{Statistical test method}
\label{sec4}
To compare the arrival direction distribution expected
from the model and that of the observed data, we need the
method by which we can represent the characteristic of the
distribution of arrival direction and apply the statistical
test easily. We proposed some comparison methods in the
previous paper \cite{Kim:2010zb,Kim:2012en}. In this analysis,
we take the correlational angular distance distribution
(CADD) method which is most appropriate for the test of
the correlation between point sources and UHECR.
CADD is the distribution of the angular distances between
all pairs of the point source and UHECR arrival directions:
\begin{equation}
\hbox{CADD : }
\left\{ \theta_{ij'}\equiv \arccos\,(\,\hat{\bf r}_i\cdot\hat{\bf r}'_j\,)
\;|\; i=1,\dots,N;\; j=1,\dots,M \right\},
\end{equation}
where $\hat{\bf r}_i$ are the UHECR arrival directions,
$\hat{\bf r}'_j$ are the point source directions,
and $N$ and $M$ are their total numbers, respectively.
Now, we get two CADDs to compare: ${\rm CADD_O}$ from
the observed UHECR and the $\gamma$-ray loud AGN,
and ${\rm CADD_M}$ from the mock UHECR of the model
under consideration and the same $\gamma$-ray loud AGN.
The total number of the data in CADD is $N_{\rm CADD}=NM$,
which means that the number of data in CADD are larger than
the sampling number N.

By comparing ${\rm CADD_O}$ and ${\rm CADD_M}$, we can test
whether our models are suitable to describe the observation.
There are several statistical test methods which can prove
that two distributions are different or not.
One of the widely used statistical
test methods is Kolmogorov-Smirnov (KS) test.
It uses the KS statistic, which is
the maximum absolute difference ($D_{\rm KS}$) between
two cumulative probability distributions (CPD), CPD of the
observed ${\rm CADD_O}$, $S_{O}(x)$, and that of the
theoretically expected ${\rm CADD_M}$, $S_{M}(x)$,
\begin{equation}
D_{\rm KS}=\max_{x}\left|S_{O}(x)-S_{M}(x)\right|.
\end{equation}
Once we calculate the KS statistic, we can get the probability
that two different distributions come from the same population
through the Monte-Carlo simulation. To get ${\rm CADD_M}$
accurately, we generates $10^5$ mock UHECR data. To obtain
the probability distribution of the KS statistic $D_{\rm KS}$,
we generate $10^5$ $D_{\rm KS}$ for a given model. Thus
our probability estimation is reliable up to roughly $10^{-4}$.

\section{Results}
\label{sec5}
In this work, we test 4 models for the UHECR sources:
1) the $\gamma$-ray loud AGN model with UHECR flux proportional to the inverse square of the distance,
2) the $\gamma$-ray loud AGN model with UHECR flux proportional to the $\gamma$-ray flux of AGN,
3) the TeV AGN model with UHECR flux proportional to the inverse square of the distance, and
4) the TeV AGN model with UHECR flux proportional to the $\gamma$-ray flux of AGN. From now on,
we call them $\gamma$-d model, $\gamma$-f model, T-d model, and T-f model, respectively.

\begin{figure}
    \begin{tabular}{cc}
        \includegraphics[width=0.5\textwidth]{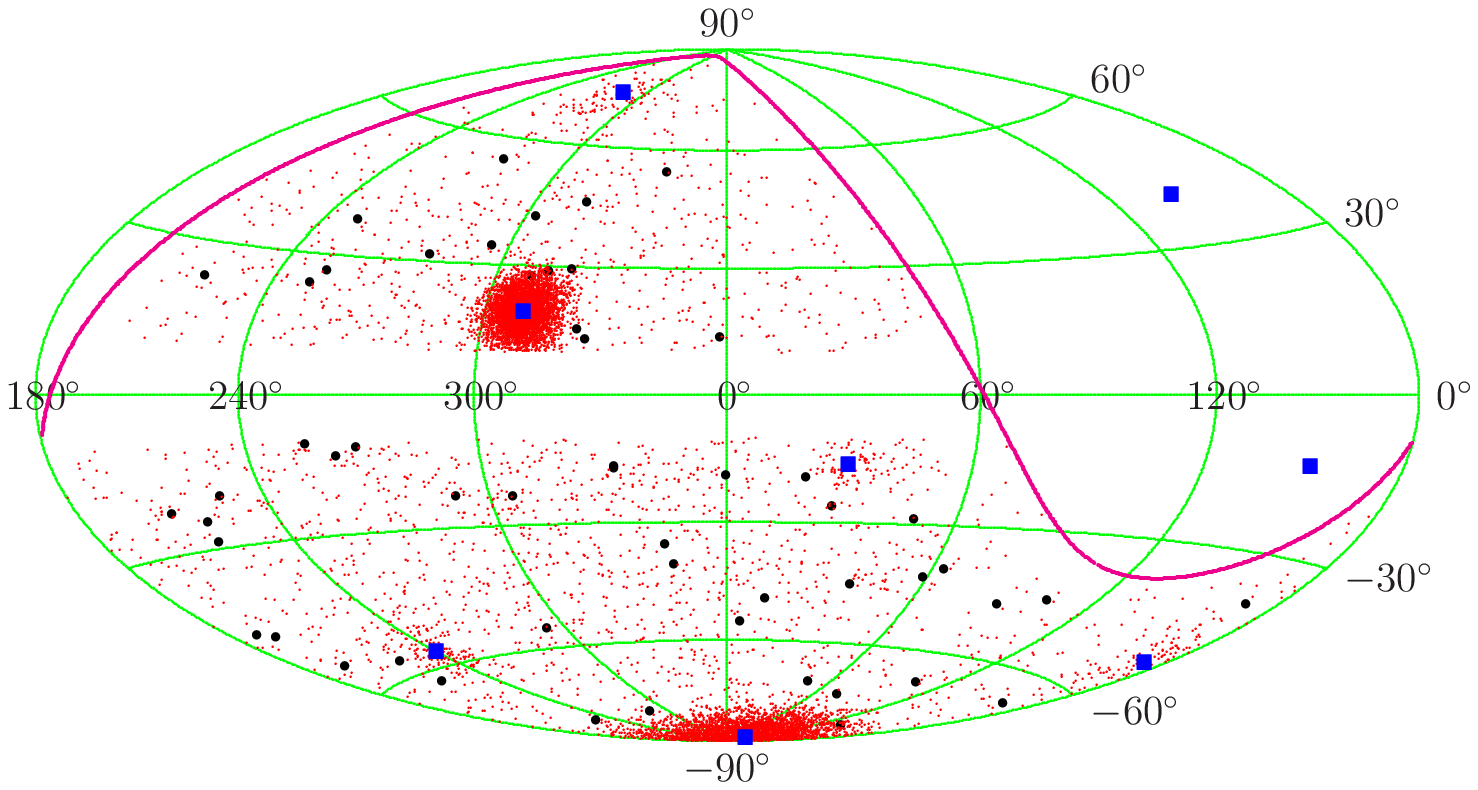} &
        \includegraphics[width=0.5\textwidth]{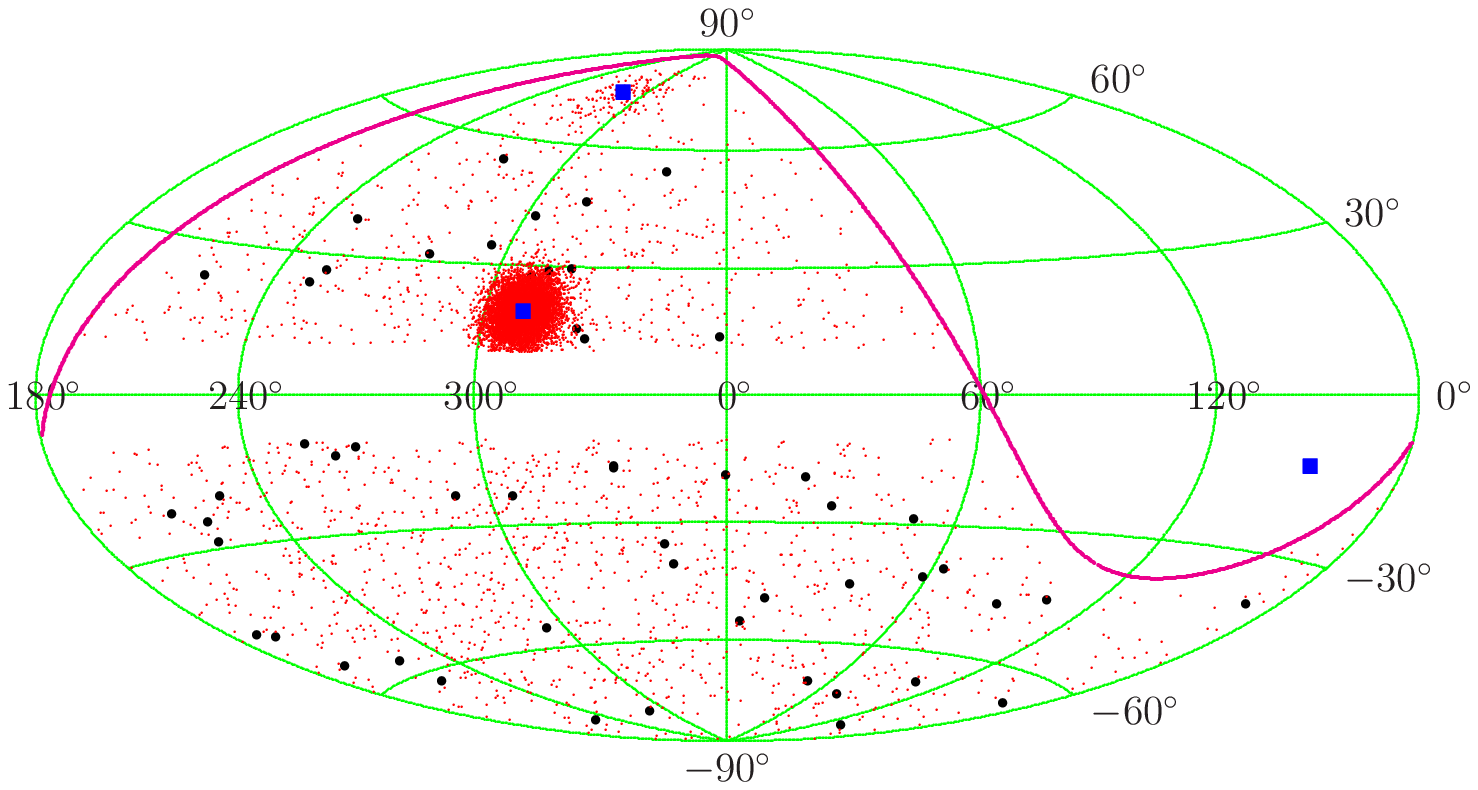}
        \\
        \includegraphics[width=0.5\textwidth]{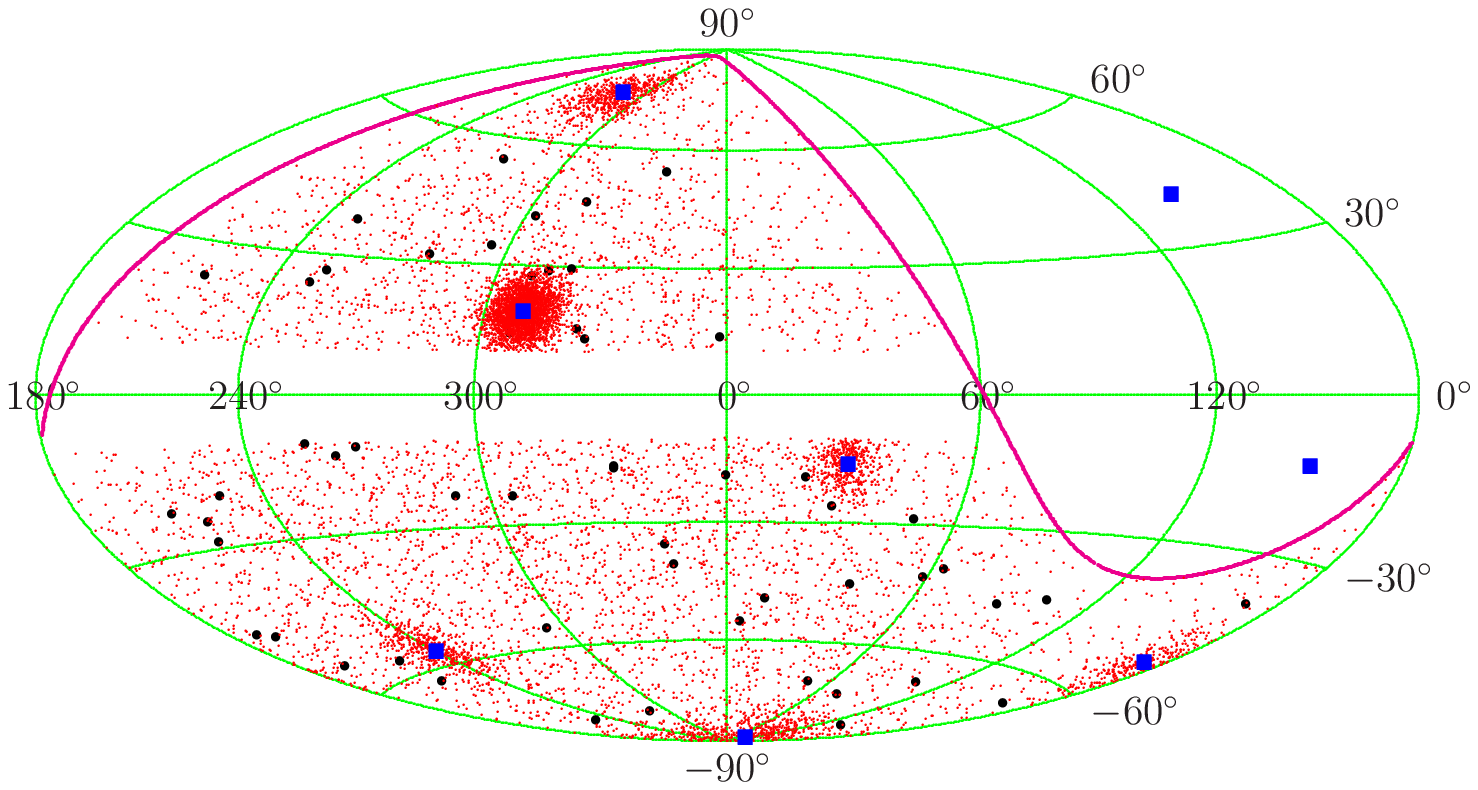} &
        \includegraphics[width=0.5\textwidth]{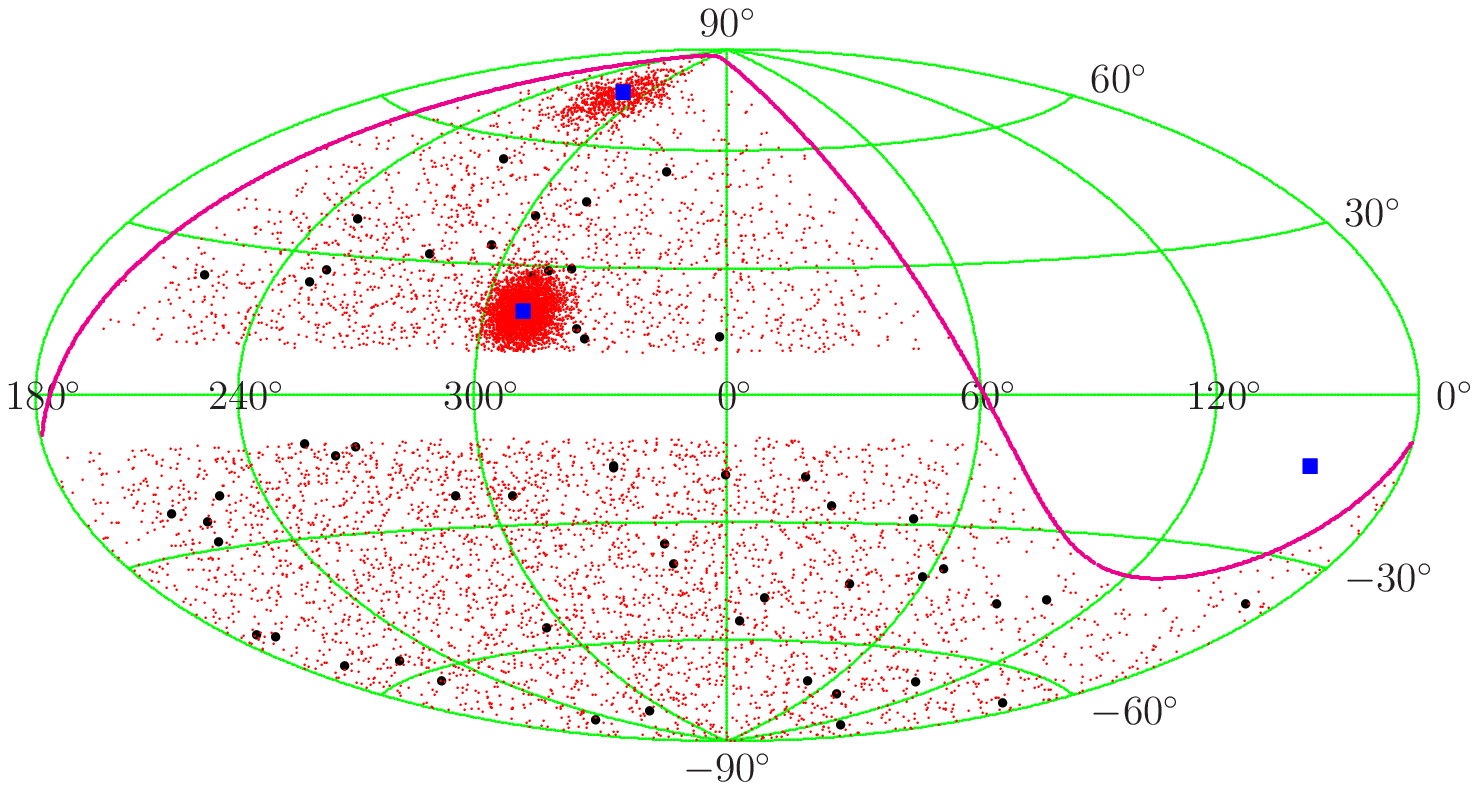}
    \end{tabular}
    \caption{
    Distributions of the mock UHECR with smearing angle $\theta_s=6^{\circ}$
    and AGN fraction $f_A = 1$ (upper panels)
    and that of the mock UHECR with AGN fraction $f_A = 0.7$ (lower panels).
    Left panels are for the $\gamma$-ray loud AGN model and
    right panels are for the TeV AGN model. The red dots are the mock
    UHECR generated by each AGN model and other marks are same
    with Fig. \ref{skymap}.
    }
\label{skymap-sim}
\end{figure}

\begin{figure}
    \begin{tabular}{cc}
        \includegraphics[width=0.5\textwidth]{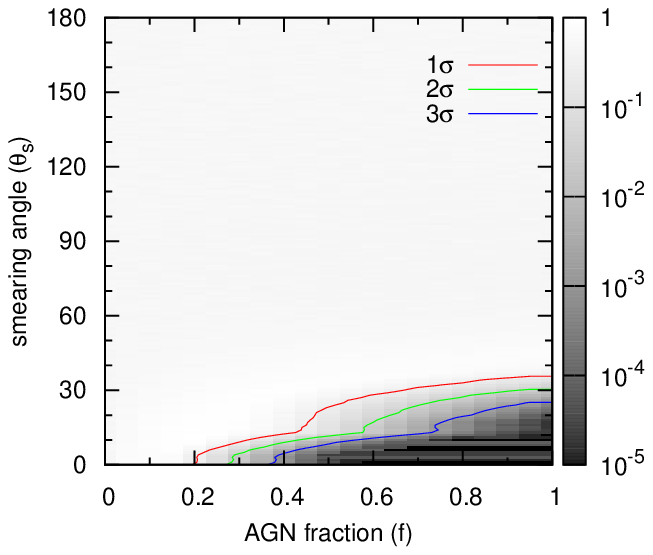} &
        \includegraphics[width=0.5\textwidth]{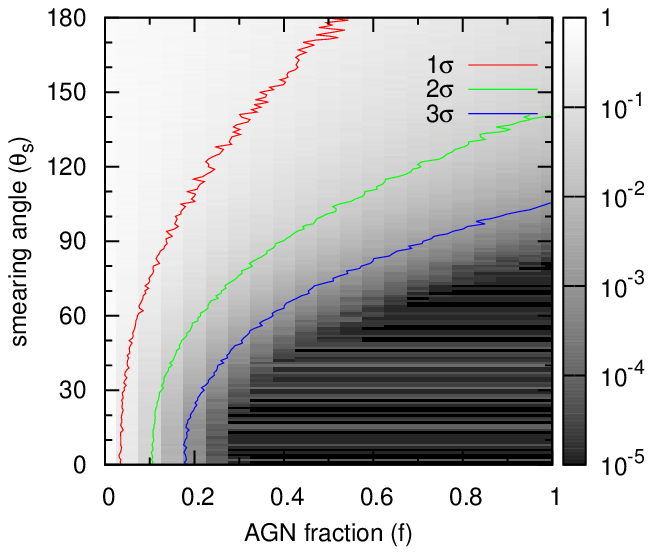}
        \\
        \includegraphics[width=0.5\textwidth]{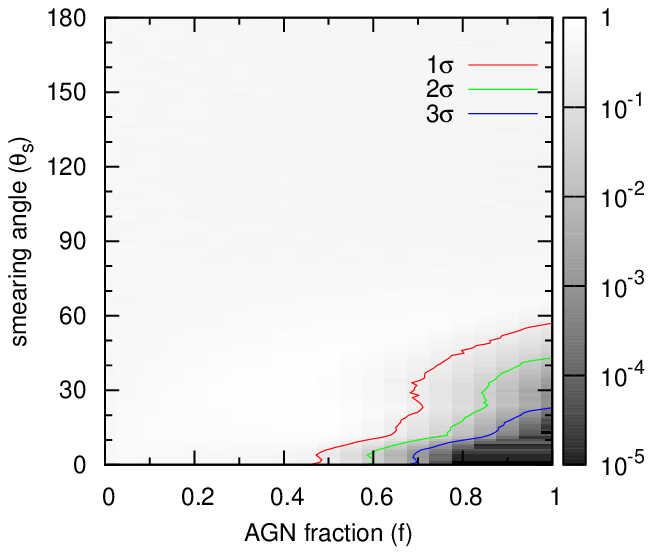} &
        \includegraphics[width=0.5\textwidth]{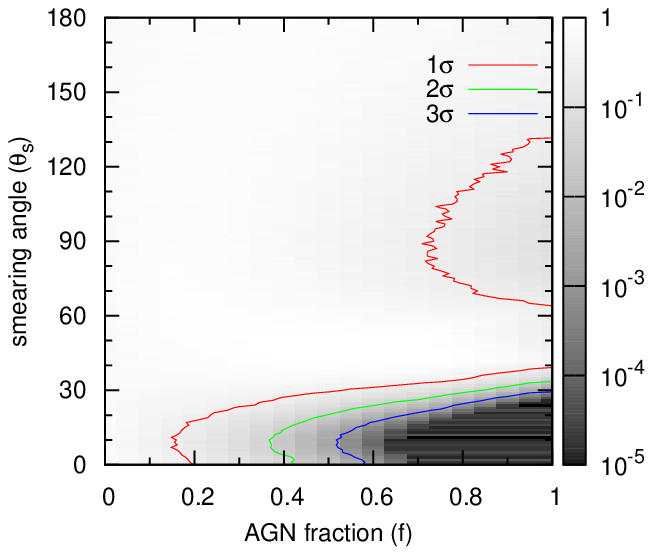}
    \end{tabular}
    \caption{
    Probability dependencies on the AGN fraction ($f_A$) and
    the smearing angle $\theta_s$. The left panel is
    for $\gamma$-ray loud AGN model and the right panel is for TeV AGN model.
    The black gradient color means the probability and the solid lines
    represent the contour plot for $1\sigma$ (red), $2\sigma$ (green), and $3\sigma$ (blue).
    }
\label{2D-prob}
\end{figure}


Fig.~\ref{skymap-sim} shows the distributions of the mock UHECR
of 4 models with the smearing angle $\theta_s = 6^{\circ}$ and
the AGN fraction $f_A = 0.7$. The two left panels are for the
$\gamma$-ray loud AGN model, i.e., $\gamma$-d model (upper panel)
and $\gamma$-f model (lower panel), and the two right panels are
for the TeV AGN model, i.e., T-d model (upper panel) and
T-f model (lower panel). The blue squares mark the locations of
$\gamma$-ray loud AGN and the red dots represent mock
UHECR generated from the source models. The mock UHECR concentrated
near the AGN are generated from the source AGN and the uniformly
distributed mock UHECR com from the isotropic background.

In Fig.~\ref{skymap-sim}, we can see the distinguished features
depending on the source models we assumed. There are
6 $\gamma$-ray loud AGN in the field of view of PAO
($\gamma$-d model and $\gamma$-f model), and there are
2 TeV AGN among them (T-d model and T-f model). Also,
we can see the difference in the mock UHECR distributions due to
the difference in UHECR flux modeling. For the $\gamma$-d model,
Cen A and NGC 0253 are the dominant sources because they are close to us.
In contrast, for the $\gamma$-f model, 6 AGN contribute rather equally to the
generation of mock UHECR. For the T-d model, Cen A is a dominant
source, but Cen A and M87 share the proportion to generate the mock
UHECR in the T-f model.

In Fig.~\ref{2D-prob}, the probability of each model as a function
of the AGN fraction from 0 to 1 and the smearing angle from $0^{\circ}$
to $180^{\circ}$ is given. The black gradient color represents the
probability that the arrival direction distribution of the PAO UHECR
comes from the given AGN source model. The red, green, and blue
contours represent $1\sigma$, $2\sigma$, and $3\sigma$ probability
lines.

First, let us compare the results of the $\gamma$-ray loud
AGN models and TeV AGN models with UHECR flux
proportional to the inverse square of the distance. Because
the set of TeV AGN is a subset of $\gamma$-ray loud AGN,
the principal difference between two models is the number of
source AGN. It brings a significant difference in the results,
which is shown in the upper panels in Fig.~\ref{2D-prob}
($\gamma$-d model and T-d model). As a rule of thumb, we choose
the $3\sigma$ probability as a criterion for ruling out the model.
The critical value of the AGN fraction is
$f_{A,c}\sim0.4$ and the critical value of the smearing
angle is $\theta_{s,c}\sim25^\circ$ for the $\gamma$-ray
loud AGN model. Decreasing the AGN fraction from the critical value
or increasing the smearing angle from the critical value
increases the probability. In comparison, for the TeV model,
the critical value of the AGN fraction is $f_{A,c}\sim0.2$
and the critical value of the smearing angle is
$\theta_{s,c}\sim110^\circ$. We can deduce that it is hard to
describe the observed UHECR distribution from T-d model.

Next, let us compare the results of two models with the same
source AGN but with different UHECR flux models: the model
which assumes that UHECR flux is proportional to the inverse
square of the distance of source AGN and the model which
assumes that UHECR flux is proportional to the $\gamma$-ray
flux of source AGN. When we compare $\gamma$-d model to
$\gamma$-f model, the critical values of the AGN fraction and
the smearing angle are $f_{A,c}\sim0.4$ and $\theta_{s,c}\sim25^\circ$
for the $\gamma$-d model, and the critical values of the AGN fraction and
the smearing angle are $f_{A,c}\sim0.7$ and $\theta_{s,c}\sim20^\circ$
for the $\gamma$-f model. We can exclude the AGN models
in the cases having the parameters within the critical values (blue line).
There is no crucial distinction between the different UHECR flux models
for the $\gamma$-ray loud AGN models.

However, when we compare T-d model to T-f model, the different UHECR
flux models result in the meaningful distinction. In the T-d model,
Cen A is a dominant source to generate the mock UHECR, thus they
are clustered around Cen A. Nevertheless, there is no significant difference
among Cen A, M87, and NGC 1275 to produce the mock UHECR in the T-f model,
thus the mock UHECR are clustered not only around Cen A but also around M87
and NGC 1275. This is the reason why the probability is increased
dramatically as the smearing angle increases in the T-f model.
The proportion of the mock UHECR produced by NGC 1275 which is located
outside of the field of view increases as the smearing angle increases.
Therefore, the clustered feature of the T-d model is quite different
from the T-f model and the small number of source cause the clear discrepancy.

In short, we find that the source models assuming $\gamma$-ray loud AGN
are responsible for UHECR ($\gamma$-d model and $\gamma$-f model)
are more plausible than the source models assuming
AGN having higher energy are responsible for UHECR (T-d model and T-f model).
Also, which flux model is appropriate for describing the UHECR flux
is not conclusive yet. This seems worthwhile to continue to study.
At this stage, we can state the critical values that the null hypotheses
are rejected. The critical regions are inside the $3\sigma$ contours.
For the $\gamma$-d model, the critical values of the AGN fraction
and the smearing angle are $f_{A,c}\sim0.4$ and $\theta_{s,c}\sim25^\circ$,
and for the $\gamma$-f model, the critical values of the AGN fraction
and the smearing angle are $f_{A,c}\sim0.7$ and $\theta_{s,c}\sim20^\circ$.
The critical values are $f_{A,c}\sim0.2$ and $\theta_{s,c}\sim110^\circ$
in the case of the T-d model and the critical values are $f_{A,c}\sim0.6$
and $\theta_{s,c}\sim30^\circ$ in the case of the T-f model. That is,
the $\gamma$-ray loud AGN dominance models with small smearing angle
are excluded.

\section{Discussion and Conclusion}
\label{sec6}
The purpose of this work was to test the possibility of a subclass
of AGN which emit strong $\gamma$-rays as the source of UHECR.
We cannot confirm that the source model using $\gamma$-ray loud AGN
is better than the source model using whole AGN in explaining the
distribution of arrival direction of UHECR. Compared to the results
of our previous work \cite{Kim:2012en}, which assumes that whole
AGN in the VCV catalog are the source of UHECR, the contour
of $1\sigma$ is changed and the overall probability distribution
for each hypothesis is higher than the previous results.
(See the probability plot for the $\gamma$-d model and
the left panel of Fig. 7 in \cite{Kim:2012en}.) However, we cannot
tell $\gamma$-ray loud AGN model describes the observation well
definitely because the higher probabilities mean that the simulated
distribution by models are consistent with the observed distribution only.

Nemmen et al. \cite{Nemmen:2010bp} investigate the correlation of UHECR
with $\gamma$-ray loud AGN using the first 27 PAO data and 1LAC.
Since the data set and the constraints used in our work and those used in their
work are different, we need to compare the results carefully.
To test cross-correlation between PAO and 1LAC, they count the
cumulative number of correlated UHECR as a function of angular
distance from the 1LAC sources located within $200\,{\rm Mpc}$.
They find that the angular distance $\psi\approx17^{\circ}$ which minimizes
the probability that the observed distribution would be the isotropic one.
Although we pointed out that this method is not suitable to test
the cross-correlation directly because it focuses on
the deviation from isotropy rather than the direct correlation
in the previous paper \cite{Kim:2010zb}, the result is consistent
with our results that $\gamma$-ray loud AGN dominance
models having small smearing angle could be rejected.

The study of arrival direction of UHECR and its source
could be one way of estimating the magnitude of the
intervening magnetic fields or identifying the primary particle.
If it is true that the $\gamma$-ray loud AGN could be the source
of UHECR, the large smearing angle may imply the large deflections
by intervening magnetic field. Interestingly, the larger deflection
angles than traditional prediction ones are estimated by
Ryu et al. \cite{Ryu:2009pf}, also. These may support the
measurements of the shower maximum by PAO \cite{Abraham:2010yv}---
the primary particle is presumed to be heavy nuclei.
In addition, the results by Dermer et al. \cite{Dermer:2010iz}
show that heavy nuclei are more likely to be
accelerated to ultra-high energy in AGN. Taken together, it is
possible that the primary particle would be heavy nuclei.

In summary, we tested the possibility that the $\gamma$-ray
emitting AGN are the sources of UHECR. We took two sub-classes
of AGN. One is the set of AGN emitting strong $\gamma$-rays
observed by Fermi LAT and the other one is the subset of
the former having $\gamma$-rays in TeV band. For the UHECR flux,
we considered two possibilities: UHECR flux is proportional
to the inverse square of the distance times the luminosity of
its source and UHECR flux is proportional to the photon flux of AGN.
We used UHECR with $E>55{\rm EeV}$ and based on GZK suppression,
we restricted AGN to be within $d_c=100\,{\rm Mpc}$.
Also, we introduced two
free parameters in the simulation to produce the mock UHECR,
AGN fraction $f_A$ and smearing angle $\theta_s$ so that
we could manipulated the contribution of the source candidate
and the deflection by intervening magnetic fields.
To compare the distributions of observed UHECR to that of
the mock UHECR, we adopted CADD which is suitable to test
cross-correlation between UHECR and sources. The probabilities
for 4 models calculated by KS test using CADD tells us that TeV AGN
model with UHECR flux proportional to the inverse square of the distance
is not appropriate to depict the observation. Rather, TeV AGN model
with UHECR flux proportional to the photon flux is more plausible
to describe the observation. In the cases of $\gamma$-ray
loud AGN models, the effects of different flux models was not
significant. If we reject the AGN models having the parameters
within the $3\sigma$ contour, the critical values
are $f_{A,t}\sim0.4$ and $\theta_{s,t}\sim25^\circ$ for the
the $\gamma$-ray loud AGN models with UHECR flux proportional to the
inverse square of the distance, and the critical values
are $f_{A,t}\sim0.7$ and $\theta_{s,t}\sim20^\circ$ for the
the $\gamma$-ray loud AGN models with UHECR flux proportional
to the $\gamma$-ray flux. This means that the models having
large isotropic background with $\gamma$-ray loud AGN or
the $\gamma$-ray loud AGN dominance models with large deflection
by the intervening magnetic fields make them possible to describe
the observed UHECR. At this stage, it is hard to confirm
that $\gamma$-ray loud AGN are the source of UHECR.

\section*{ACKNOWLEDGMENT}
This work was supported by the research fund of Hanyang University (HY-2006-S).

\end{document}